\documentclass[11pt]{article}
\usepackage{a4,bm}
\newcommand{\laenge}{\ell}
\newcommand{\qm}{m}
\newcommand{\qqm}{{\quer m}}
\newcommand{\vorz}{-}

\newcommand{\qquer}{}
\newcommand{\ph}[1]{\phantom{#1}}
\newcommand{\aniso}{{\text{aniso}}}
\newcommand{\Bohr}{{\mathrm{Bohr}}}
\newcommand{\rb}{\overline\R_\Bohr}
\newcommand{\bpm}{\begin{pmatrix}}
\newcommand{\epm}{\end{pmatrix}}
\newcommand{\sq}{\Delta}
\newcommand{\vc}{{\mathbf{c}}}
\newcommand{\fff}{f}
\newcommand{\invzush}{A_0}
\renewcommand{\invzush}{A_{\bullet}}
\renewcommand{\invzush}{A_{\ast}}
\newcommand{\ffa}{\rho}
\newcommand{\ffb}{\sigma}
\newcommand{\ffh}{\varphi}
\newcommand{\sqausgeschrieben}{\sqrt{\frac{\lambda^2}4+ c(c-\varkappa)}}
\newcommand{\sqandersausgeschrieben}{\sqrt{\Bigl(c - \frac\varkappa2\Bigr)^2 + \frac{\lambda^2-\varkappa^2}4}}

\newcommand{\quasienergie}{{\cal E}}

\newcommand{\Bigogkl}[1]{\Bigl\lceil#1\Bigr\rceil}
\newcommand{\fktabgesetzt}[3]%                %Fast zentriert in zwei extra Zeilen
           {
           
            \hspace*{\fill}
            $\begin{array}[t]{cccc}%
            #1: & #2 & \nach & #3 
            \end{array}$
            \settowidth{\fktdefhilfslaenge}{$#1$:}
            \hspace*{0.6 \fktdefhilfslaenge}  
            \hspace*{\fill}
            
            }
\usepackage{ifthen,array}
\newcommand{\ams}{\usepackage{amsfonts,amssymb,amsmath}}

\ams
\allowdisplaybreaks[3]
\newlength{\textwidthorig}
\newlength{\oddsidemarginorig}
\newlength{\textheightorig}
\newlength{\topmarginorig}
\def\seitenlaengenabsolut#1 #2 #3 #4 {\setlength{\textwidth}{#1}
                                      \setlength{\oddsidemargin}{#2}
                                      \setlength{\textheight}{#3}
                                      \setlength{\topmargin}{#4}}
\def\seitenlaengenrelzustandard#1 #2 #3 #4 {\setlength{\textwidth}{\textwidthorig+#1}
                                            \setlength{\oddsidemargin}{\oddsidemarginorig+#2}
                                            \setlength{\textheight}{\textheightorig+#3}
                                            \setlength{\topmargin}{\topmarginorig+#4}}
\def\seitenlaengenrelzuvorher#1 #2 #3 #4 {\addtolength{\textwidth}{#1}
                                          \addtolength{\oddsidemargin}{#2}
                                          \addtolength{\textheight}{#3}
                                          \addtolength{\topmargin}{#4}}
\newcommand{\standardseite}{\seitenlaengenrelzuvorher2.2cm -0.8cm 1.8cm -1.5cm }   %
\standardseite
\newcommand{\leerezeile}{\vspace{2ex}}
\newlength{\laengespatium}

\newcommand{\nach}{\longrightarrow}      %Abbildungspfeil

\newcommand{\auf}{\longmapsto}           %Abbildungspfeil fuer Elemente
\newcommand{\txtauf}[1]{\auf}            %Abbildungspfeil mit Opt. "Text dr"uber"
\newcommand{\impliz}{\Longrightarrow}    %Implikationspfeil
\newcommand{\invimpliz}{\Longleftarrow}  %Implikationspfeil (umgekehrte Richtung
\newcommand{\gegen}{\rightarrow}         %Konvergenzpfeil
\newcommand{\ident}{\equiv}              %Symbol fuer identisch (3 Striche)
\newcommand{\teilmenge}{\subseteq}       %Symbol fuer Teilmenge
\newcommand{\obermenge}{\supseteq}       %Symbol fuer Teilmenge andersherum
\newcommand{\echteobermenge}{\supset}    %Symbol fuer echte Teilmenge andersherum
\newcommand{\aeqrel}{\sim}               %Symbol Aequivalenzrelation
\newcommand{\kreuz}{\times}              %Symbol f"ur Kreuz

\newcommand{\betraganpass}[1]%
           {\left| #1 \right|}           %Betragsstriche (variable Gr"o"se) 
\newcommand{\bigbetrag}[1]%
           {\bigl|{#1}\bigr|}            %Betragsstriche (gr"o"ser)
\newcommand{\betrag}[1]%
           {|{#1}|}                      %Betragsstriche  
\newcommand{\betragnichtanpass}[1]%
           {\mid #1 \mid}                %Betragsstriche 
\newcommand{\norm}[1]%
           {{}{\parallel}#1{\parallel}{}}      %Normstriche  
\newcommand{\erww}[1]%
           {\langle #1 \rangle}          %Erwartungwert-Klammern
\newcommand{\skalprod}[2]%
           {\langle #1,#2 \rangle}       %eckiges Skalarprodukt
\newcommand{\supnorm}[1]{{\norm{#1}_\infty}}        %Supremumsnorm allgemein
\newcommand{\quer}{\overline}            %Strich drueber
\newcommand{\inv}[1]{\frac{1}{#1}}       %Liefert 1/#1 als Bruch
\newcommand{\re}{\text{Re }}                           %Re als Realteil
\newcommand{\diag}{\text{diag }}                       %diag als Diagonalmatrix
\newcommand{\del}{\partial}                            %krummes d fuer partielle Abl.
\newcommand{\Hom}{\text{Hom}}                          %Hom als Homomorphismengruppe
\newcommand{\dd}{\text{d}}                             %Differential d
\newcommand{\e}{\text{e}}                              %exp
\newcommand{\I}{\text{i}}                              %i als imagin"are Einheit
\newcommand{\field}[1]{\mathbb{#1}}                    %liefert #1 als mathbb-Zeichen
\newcommand{\C}{{\field{C}}}                           %C fuer komplexe Zahlen
\newcommand{\N}{{\field{N}}}                           %N fuer natuerliche Zahlen
\newcommand{\R}{{\field{R}}}                           %R fuer reelle Zahlen
\newcommand{\rnkl}[2]{\raisebox{-0.4ex}{$#1$}%
\raisebox{-0.12ex}{{\large$\setminus$}}\,#2}   %0.5/0.2 fr"uher  %Rechtsnebenklassenfaktorraum
\newcommand{\agb}{{\overline{{\cal A}/{\cal G}}}}      %A/G + Strich
\newcommand{\agbfact}[1][]{\text{$\agb/\!\aeqrel$}}    %A/G\~ + Strich
\newcommand{\Ab}{{\overline{{\cal A}}}}                %A + Strich
\newcommand{\A}{{\cal A}}                              %A ohne Strich
\newcommand{\Gb}{{\overline{{\cal G}}}}                %G + Strich

\newcommand{\qa}{{\quer{A}}}                           %A als verallg. Zusammenhang 
\newcommand{\holgr}{{\mathbf H}}                       %Holonomiegruppe
\newcommand{\bz}{{\mathbf B}}                          %Basiszentralisator
\newcommand{\Pf}{{\cal P}}                             %Menge aller Pfade
\newcommand{\LG}{{\mathbf{G}}}                         %Liegruppe G (fett)
\newcommand{\Lieg}{{\mathfrak{g}}}                            %gotisches g (fuer Liealgebra)
\newcommand{\aeqrelzush}[1][]{\sim}                    %Symbol Aequivalenzrelation
\newcommand{\nklza}[1][]{\ifthenelse{\equal{#1}{}}     %Z(H_\qa) \ G
                                    {\rnkl{Z(\holgr_\qa)}{\LG}}        
                                   {\rnkl{Z(\holgr_{#1})}{\LG}}}       
\newcommand{\nkla}[1][]{\ifthenelse{\equal{#1}{}}      %B(\qa) \ \Gb
                                    {\rnkl{\bz(\qa)}{\Gb}}        
                                    {\rnkl{\bz(#1)}{\Gb}}}       
\newcommand{\YM}{{\text{YM}}}                          %YM f"ur Yang-Mills

\newcommand{\ymwirk}[1][]{\ifthenelse{\equal{#1}{}}{S_{\YM}}{S_{\YM,#1}}}
\newcommand{\bmat}{\begin{pmatrix}}
\newcommand{\emat}{\end{pmatrix}}

\newcommand{\ListNullAbstaende}{\setlength{\topsep}{0pt}%
                                \setlength{\parskip}{0pt}%
                                \setlength{\partopsep}{0pt}%
                                \setlength{\itemsep}{0pt}%
                                \setlength{\parsep}{0pt}}
\newcommand{\ListNurAnstrichAbstand}{\setlength{\topsep}{0pt}%
                                     \setlength{\parskip}{0pt}%
                                     \setlength{\partopsep}{0pt}%
                                     \setlength{\parsep}{0pt}}
\newenvironment{StandardListe}[2]%
               {\begin{list}%
                      {#1}%
                      {\settowidth{\leftmargin}{M#1}%
                       \settowidth{\labelwidth}{#1}%
                       \settowidth{\labelsep}{M}%
                       #2%
                      }%
                }%
               {\end{list}}%
\newenvironment{EinfachListe}[1]%
               {\begin{StandardListe}{#1}{\ListNullAbstaende}}%
               {\end{StandardListe}}%
               {\begin{StandardListe}{#1}{\ListNurAnstrichAbstand}}%
               {\end{StandardListe}}%
\newcommand{\labelsatz}[1]{#1}
\newcounter{listennr}                      %
\newlength{\hilfslaenge}
\newlength{\stdlabellaenge}
\newlength{\maximum}
\newcommand{\stdlabel}{}
\newcommand{\Maximum}{}
\newcommand{\iitem}[1][]{\ifthenelse{\equal{#1}{}}%
                           {\item \setlength{\hilfslaenge}{\stdlabellaenge}}%
                           {\item[\labelsatz{#1}\hfill]%
                            \settowidth{\hilfslaenge}{\labelsatz{#1}}}%
                         \ifthenelse{\lengthtest{\maximum < \hilfslaenge}}%
                           {\setlength{\maximum}{\hilfslaenge}%
                            \ifthenelse{\equal{#1}{}}%
                               {\renewcommand{\Maximum}{\stdlabel}}%
                               {\renewcommand{\Maximum}{#1}}}%
                           {}%
                      }      
\makeatletter
\newenvironment{AutoLabelLaengenListe}[2][]%
               {\begin{list}%
                      {\labelsatz{#1}\hfill}%
                      {\stepcounter{listennr}%
                       \settowidth{\leftmargin}{M\labelsatz{\ref{listnr\arabic{listennr}}}}%
                       \settowidth{\labelwidth}{\labelsatz{\ref{listnr\arabic{listennr}}}}%
                       \settowidth{\labelsep}{M}%
                       \settowidth{\stdlabellaenge}{\labelsatz{#1}}%
                       \renewcommand{\stdlabel}{#1}%
                       #2%
                       \renewcommand{\Maximum}{}%
                      }%
                }%
               {\renewcommand{\@currentlabel}{\Maximum}%
                \label{listnr\arabic{listennr}}%
                \end{list}%
                }%
\makeatother
\newenvironment{StandardEinrueckung}[2]%
               {\begin{list}%
                      {#1}%
                      {\settowidth{\leftmargin}{M#1}%
                       \settowidth{\labelwidth}{#1}%
                       \settowidth{\labelsep}{M}%
                       #2%
                      }%
                \item}%
               {\end{list}}%
\newenvironment{Einrueckungpur}[1]%
               {\begin{StandardEinrueckung}{#1}{\ListNullAbstaende}}%
               {\end{StandardEinrueckung}}%
\newenvironment{Einrueckung}[1]%
               {\begin{StandardEinrueckung}{#1}{\setlength{\parsep}{0pt}}}%
               {\end{StandardEinrueckung}}%

\makeatletter
\newcommand{\EineNumZeileGleichung}[2][0.5ex]
           {
            
            \vspace{#1} 
            \noindent
            \stepcounter{equation}
            \renewcommand{\@currentlabel}{\arabic{equation}}%
            \phantom{(\arabic{equation})}\hspace*{\fill}
            $\displaystyle{#2}$
            \hspace*{\fill}
            (\arabic{equation})

            \vspace{#1} 
            
           }
\makeatother
\makeatletter
\newcommand{\EineErwNumZeileGleichung}[2][0.5ex]
           {
            
            \vspace{#1} 
            \noindent
            \stepcounter{equation}
            \renewcommand{\@currentlabel}{\arabic{equation}}%
            \phantom{(\arabic{equation})}\hspace*{\fill}
            #2 %
            \hspace*{\fill}
            (\arabic{equation})

            \vspace{#1} 
            
           }
\makeatother
\newcommand{\breitrel}[1]{\hspace*{\tabcolsep} #1 \hspace*{\tabcolsep}}
\newlength{\abstaug}              %
\newenvironment{AllgUnnumGleichung}[2][1.0ex]%           %
               {
  
                \setlength{\abstaug}{#1}
                \vspace{\abstaug}
                \hspace*{\fill}
                $\begin{array}[t]{#2}
                }%
               {\end{array}$
                \hspace*{\fill}
  
                \vspace{\abstaug}

                }%
\newenvironment{AllgNumGleichung}[2][0.0ex]%           %
               {
  
                \setlength{\abstaug}{#1}
                \vspace{\abstaug}
                $\begin{tabular*}{\textwidth}[t]{#2}
                }%
               {\end{tabular*}$

                \vspace{\abstaug}

               }%
\newenvironment{StandardUnnumGleichungKlein}[1][0ex]%       %
               {\renewcommand{\s}{\\[#1] }%
                \begin{AllgUnnumGleichung}{rcl}}%
               {\end{AllgUnnumGleichung}}%
\newcommand{\s}{\\[0ex] }             %
\newenvironment{StandardUnnumGleichung}[1][0ex]%       
               {\renewcommand{\s}{\\[#1] }%
                \begin{AllgUnnumGleichung}{>{\displaystyle}rc>{\displaystyle}l}}%
               {\end{AllgUnnumGleichung}}%
\newenvironment{XrelYZNumGleichung}[1][0ex]%       %
               {\renewcommand{\s}{\\[#1] }%
                \begin{AllgNumGleichung}{rcll}}%
               {\end{AllgNumGleichung}}%
\newcommand{\erl}[1]{\hfill\mbox{\hspace*{1.5em}\small (#1)}}

\newcommand{\erllang}[2][0.5\textwidth]%
              {\hfill\hspace*{1.5em}%
               \begin{minipage}[t]{#1}{\small%
                          \begin{list}{(}{\ListNullAbstaende%
                                          \settowidth{\leftmargin}{(}%
                                          \settowidth{\labelwidth}{(}%
                                          \settowidth{\labelsep}{}%
                                         }%
                          \item#2)%
                          \end{list}}%
               \end{minipage}\\[-0.9ex]
              }%         
\newcommand{\DefBemUmgeb}[1]% 
           {\newenvironment{#1}[1][]%
                           {\begin{Einrueckung}{{\bf #1}}%
                            \ifx##1\empty\else{{\bf ##1}
                            
                                                        }\fi%
                            }%
                           {\end{Einrueckung}}}
\newcommand{\DefSBemUmgeb}[2]% %
           {\newenvironment{#1}[1][]%
                           {\begin{Einrueckung}{{\bf #2}}%
                            \ifx##1\empty\else{{\bf ##1}
                            
                                                        }\fi%
                            }%
                           {\end{Einrueckung}}}
\makeatletter
\newcommand{\DefBspUmgeb}[3]% %
           {\newcounter{#2}[#3]%
            \newenvironment{#1}[1][]%
                           {\stepcounter{#2}%
                            \renewcommand{\ZaehlerMarke}{\arabic{#2}}%  
                            \renewcommand{\Einzugsname}{{\bf #1 \ZaehlerMarke}}%
                            \begin{Einrueckung}{\Einzugsname}
                            \ifx##1\empty\else{{\bf ##1}\\}\fi%
                            \renewcommand{\@currentlabel}{\ZaehlerMarke}%
                            }%
                           {\end{Einrueckung}}}
\makeatother
\newcommand{\ZaehlerbisEbene}{section}
\newcommand{\Ebenea}{section}
\newcommand{\Ebeneb}{subsection}

\newcommand{\Abschnittnummer}{%
            \ifx\ZaehlerbisEbene\Ebenea{\arabic{section}}%
             \else{%
              \ifx\ZaehlerbisEbene\Ebeneb{\arabic{section}.\arabic{subsection}}%
               \else{\arabic{section}.\arabic{subsection}.\arabic{subsubsection}}%
              \fi}%     
            \fi}     
\newcommand{\Abschnittnummerpunkt}{\Abschnittnummer.}     %  {}%
\newcommand{\Einzugsname}{}
\newcommand{\ZaehlerMarke}{}
\makeatletter
\newcommand{\DefThmUmgeb}[3]% 
           {\newcounter{#1}[#3]%
            \newenvironment{#1}[1][]%
                           {\stepcounter{#2}%
                            \setcounter{#1}{\value{#2}}%
                            \renewcommand{\ZaehlerMarke}{\Abschnittnummerpunkt\arabic{#1}}%  
                            \renewcommand{\Einzugsname}{{\bf #1 \ZaehlerMarke}}%
                            \begin{Einrueckung}{\Einzugsname}
                            \ifx##1\empty\else{{\bf ##1}
                            
                                                        }\fi%
                            \renewcommand{\@currentlabel}{\ZaehlerMarke}%
                            }%
                           {\end{Einrueckung}}}
\makeatother
\makeatletter
\newcommand{\DefSThmUmgeb}[4]% 
           {\newcounter{#1}[#3]%
            \newenvironment{#1}[1][]%
                           {\stepcounter{#2}%
                            \setcounter{#1}{\value{#2}}%
                            \renewcommand{\ZaehlerMarke}{\Abschnittnummerpunkt\arabic{#1}}%
                            \renewcommand{\Einzugsname}{{\bf #4 \ZaehlerMarke}}
                            \begin{Einrueckung}{\Einzugsname}
                            \ifx##1\empty\else{{\bf ##1}

                                                        }\fi%
                            \renewcommand{\@currentlabel}{\ZaehlerMarke}%
                            }%
                           {\end{Einrueckung}}}
\makeatother
\makeatletter
\newcommand{\DefUnterNumThmUmgeb}[5]% 
           {\newcounter{#1}[#3]%
            \newcounter{#4}%
            \newenvironment{#1}[1][]%
                           {\ifx##1\empty\else{\stepcounter{#2}\setcounter{#4}{0}}\fi%
                            \stepcounter{#4}%
                            \setcounter{#1}{\value{#2}}%
                            \renewcommand{\ZaehlerMarke}{\Abschnittnummerpunkt\arabic{#1}\alph{#4}}%
                            \renewcommand{\Einzugsname}{{\bf #5 \ZaehlerMarke}}
                            \begin{Einrueckung}{\Einzugsname}
                            \renewcommand{\@currentlabel}{\ZaehlerMarke}%
                            }%
                           {\end{Einrueckung}}}
\makeatother
\newenvironment{Beweis}[1][]%
               {\begin{Einrueckung}{{\bf Beweis}}%
                \ifx#1\empty\else{{\bf #1}

                                            }\fi%
                }%
               {\end{Einrueckung}%
                }%
\newenvironment{Proof}[1][]%
               {\begin{Einrueckung}{{\bf Proof}}%
                \ifx#1\empty\else{{\bf #1}

                                            }\fi%
                }%
               {\end{Einrueckung}%
                }%
               {\begin{Einrueckung}{{\bf \glqq Beweis\grqq}}%
                \ifx#1\empty\else{{\bf #1}
                
                                            }\fi%
                }%
               {\end{Einrueckung}%
                }%
               {\begin{Einrueckung}{{\bf Begr"undung}}%
                \ifx#1\empty\else{{\bf #1}
                
                                            }\fi%
                }%
               {\end{Einrueckung}%
                }%
\newenvironment{Hinrichtung}%
               {\begin{Einrueckungpur}{$\impliz$}}%
               {\end{Einrueckungpur}}%
\newenvironment{Rueckrichtung}%
               {\begin{Einrueckungpur}{$\invimpliz$}}%
               {\end{Einrueckungpur}}%
               {\begin{Einrueckungpur}{\glqq$\teilmenge$\grqq}}%
               {\end{Einrueckungpur}}%
               {\begin{Einrueckungpur}{\glqq$\obermenge$\grqq}}%
               {\end{Einrueckungpur}}%
               {\begin{Einrueckungpur}{"$\teilmenge$"}}%
               {\end{Einrueckungpur}}%
               {\begin{Einrueckungpur}{"$\obermenge$"}}%
               {\end{Einrueckungpur}}%
\newcommand{\qed}{\nopagebreak\hspace*{2em}\hspace*{\fill}{\bf qed}}
\newcommand{\ARabic}{\arabic}
\newcommand{\Nummerntypa}{\arabic}   
\newcommand{\Nummerntypb}{\alph}
\newcommand{\Nummerntypc}{\roman}
\newcommand{\Nummerntypd}{\Alph}

\newcommand{\Nra}{\Nummerntypa{Nummera}}            %
\newcommand{\Nrb}{\Nummerntypb{Nummerb}}            %
\newcommand{\Nrc}{\Nummerntypc{Nummerc}}                
\newcommand{\Nrd}{\Nummerntypd{Nummerd}}                
\newcommand{\ZeichenzuNrTyp}[1]%
           {\ifx#1\ARabic {.}\else{)}%
                  \fi}                              %
\newcommand{\NrZeicha}{\ZeichenzuNrTyp{\Nummerntypa}}
\newcommand{\NrZeichb}{\ZeichenzuNrTyp{\Nummerntypb}}
\newcommand{\NrZeichc}{\ZeichenzuNrTyp{\Nummerntypc}}
\newcommand{\NrZeichd}{\ZeichenzuNrTyp{\Nummerntypd}}
\newcommand{\ListMarkea}%
           {\Nra\NrZeicha}
\newcommand{\ListMarkeb}%
           {\Nra\NrZeicha\Nrb\NrZeichb}
\newcommand{\ListMarkec}%
           {\Nra\NrZeicha\Nrb\NrZeichb\Nrc\NrZeichc}
\newcommand{\ListMarked}%
           {\Nra\NrZeicha\Nrb\NrZeichb\Nrc\NrZeichc\Nrd\NrZeichd}
\newcommand{\Anfangszeichen}{}
\newcommand{\Anfangspunkt}{}
\newcounter{Schachtelebene}
\newcounter{Hilfszaehler}
\newcommand{\Hilfsbefehl}{}
\newcommand{\Schachtelebene}{\alph{Schachtelebene}}
\makeatletter
\newenvironment{AllgNumerierteListe}[2][]%      %
               {\addtocounter{Schachtelebene}{1}%
		\setcounter{Hilfszaehler}{#2}%
                \renewcommand{\Anfangszeichen}%
                             {\renewcommand{\Hilfsbefehl}{\csname Nummerntyp\Schachtelebene \endcsname}%
                              \Hilfsbefehl{Hilfszaehler}}%
                \renewcommand{\Anfangspunkt}%
                             {\csname NrZeich\Schachtelebene \endcsname}%
                \begin{list}%
                      {\stepcounter{Nummer\Schachtelebene}%
                       \csname Nr\Schachtelebene \endcsname
                       \csname NrZeich\Schachtelebene \endcsname
                       }%
                      {\settowidth{\leftmargin}{M\Anfangszeichen\Anfangspunkt}%
                       \settowidth{\labelwidth}{\Anfangszeichen\Anfangspunkt}%
                       \settowidth{\labelsep}{M}%
                       \setlength{\topsep}{0pt}%
                       \setlength{\parskip}{0pt}%
                       \setlength{\partopsep}{0pt}%
                       \setlength{\itemsep}{0pt}%
                       \setlength{\parsep}{0pt}%
                      }%
                \renewcommand{\@currentlabel}{\csname ListMarke\Schachtelebene \endcsname}%
                }%      
               {\ifthenelse{\equal{}{}}{\setcounter{Nummer\Schachtelebene}{0}}{}
                \addtocounter{Schachtelebene}{-1}%
                \end{list}}
\makeatother
\newenvironment{NumerierteListe}[1]%      %
               {\begin{AllgNumerierteListe}{#1}}
               {\end{AllgNumerierteListe}}
\newenvironment{WeiterNumerierteListe}[1]%      %
               {\begin{AllgNumerierteListe}[Weiter]{#1}}
               {\end{AllgNumerierteListe}}

\newcommand{\UnnumAnfangszeichen}{}
\newcounter{UnnumSchachtelebene}
\newcommand{\UnnumSchachtelebene}{\alph{UnnumSchachtelebene}}
\makeatletter
\newenvironment{UnnumerierteListe}%          
               {\addtocounter{UnnumSchachtelebene}{1}%
                \renewcommand{\UnnumAnfangszeichen}%
                             {\csname UnnumZeich\UnnumSchachtelebene \endcsname}%
                \begin{list}%
                      {\UnnumAnfangszeichen}%
                      {\settowidth{\leftmargin}{M\UnnumAnfangszeichen}%
                       \settowidth{\labelwidth}{\UnnumAnfangszeichen}%
                       \settowidth{\labelsep}{M}%
                       \setlength{\topsep}{0pt}%
                       \setlength{\parskip}{0pt}%
                       \setlength{\partopsep}{0pt}%
                       \setlength{\itemsep}{0pt}%
                       \setlength{\parsep}{0pt}%
                      }%
                }%
               {\addtocounter{UnnumSchachtelebene}{-1}%
                \end{list}}
\makeatother
\newlength{\fktdefhilfslaenge}
\newcommand{\ohnefktdef}[4]%                 %
           {\hspace*{\fill}
            $\begin{array}[t]{ccc}%
            #1 & \nach & #2 \\
            #3 & \auf  & #4
            \end{array}$
            \hspace*{\fill}}
\newcommand{\fktdef}[5]%                 %
           {\hspace*{\fill}
            $\begin{array}[t]{cccc}%
            #1: & #2 & \nach & #3 \\    
                & #4 & \auf  & #5
            \end{array}$
            \settowidth{\fktdefhilfslaenge}{$#1$:}
            \hspace*{0.6 \fktdefhilfslaenge}  
            \hspace*{\fill}}
\newcommand{\fktdefpur}[5]%                 %
           {$\begin{array}[t]{cccc}%
            #1: & #2 & \nach & #3 \\    
                & #4 & \auf  & #5
            \end{array}$}
\newcommand{\fktdefabgesetztpur}[5]%          %
           {
            
            $\begin{array}[t]{cccc}%
            #1: & #2 & \nach & #3 \\    
                & #4 & \auf  & #5
            \end{array}$
            \settowidth{\fktdefhilfslaenge}{$#1$:}
            \hspace*{0.6 \fktdefhilfslaenge}
            
           }
\newcommand{\fktdefabgesetzt}[5]%                %
           {
           
            \hspace*{\fill}
            $\begin{array}[t]{cccc}%
            #1: & #2 & \nach & #3 \\    
                & #4 & \auf  & #5
            \end{array}$
            \settowidth{\fktdefhilfslaenge}{$#1$:}
            \hspace*{0.6 \fktdefhilfslaenge}  
            \hspace*{\fill}
            
            }
\newcommand{\ohnefktdefabgesetzt}[4]%                %
           {      

            \hspace*{\fill}
            $\begin{array}[t]{ccc}%
            #1 & \nach & #2 \\
            #3 & \auf  & #4
            \end{array}$
            \hspace*{\fill}

            }
\newcommand{\doppelohnefktdefabgesetzt}[6]%                %
           {

            \hspace*{\fill}
            $\begin{array}[t]{ccccc}%
            #1 & \nach & #2 & \nach & #3\\
            #4 & \auf  & #5 & \auf  & #6
            \end{array}$
            \hspace*{\fill}

            }
\newcommand{\anhang}%
           {\appendix
            \sectioninh{Anhang}
            \renewcommand{\Abschnittnummer}{%
                  \ifx\ZaehlerbisEbene\Ebenea{\Alph{section}}%
                  \else{%
                        \ifx\ZaehlerbisEbene\Ebeneb{\Alph{section}.\arabic{subsection}}%
                        \else{\Alph{section}.\arabic{subsection}.\arabic{subsubsection}}%
                        \fi}%     
                  \fi}%
            \renewcommand{\Abschnittnummerpunkt}{\Abschnittnummer.}     
            }            
\newcommand{\anhangengl}%
           {\appendix
            \sectioninh{Appendix}
            \renewcommand{\Abschnittnummer}{%
                  \ifx\ZaehlerbisEbene\Ebenea{\Alph{section}}%
                  \else{%
                        \ifx\ZaehlerbisEbene\Ebeneb{\Alph{section}.\arabic{subsection}}%
                        \else{\Alph{section}.\arabic{subsection}.\arabic{subsubsection}}%
                        \fi}%     
                  \fi}%
            \renewcommand{\Abschnittnummerpunkt}{\Abschnittnummer.}     
            }

\newcounter{wdhlstufe}
\newcommand{\sectioninh}[1]%
           {\section*{#1}%
            \addcontentsline{toc}{section}{#1}}
\newcommand{\bezeichnung}[3]%
           {\begin{Einrueckungpur}{\hbox to 6em{#1}\hbox to 2.4em{\hfill#2}}
            #3
            \end{Einrueckungpur}}

\newcommand{\doppelteinfach}{e}

\newcommand{\ifdoppelt}[1]{\ifthenelse{\equal{\doppelteinfach}{d}}{#1}{}}
\newcommand{\ifeinfach}[1]{\ifthenelse{\equal{\doppelteinfach}{e}}{#1}{}}

\newlength{\querfhilfsl}              %

\newlength{\hll}

\DefThmUmgeb{Theorem}{Theorem}{\ZaehlerbisEbene}
\DefThmUmgeb{Definition}{Definition}{\ZaehlerbisEbene}
\DefThmUmgeb{Satz}{Theorem}{\ZaehlerbisEbene}
\DefThmUmgeb{Proposition}{Theorem}{\ZaehlerbisEbene}
\DefThmUmgeb{Lemma}{Theorem}{\ZaehlerbisEbene}
\DefThmUmgeb{Folgerung}{Theorem}{\ZaehlerbisEbene}
\DefThmUmgeb{Corollary}{Theorem}{\ZaehlerbisEbene}
\DefThmUmgeb{Vorschrift}{Definition}{\ZaehlerbisEbene}
\DefThmUmgeb{Construction}{Definition}{\ZaehlerbisEbene}
\DefSThmUmgeb{FormSatz}{Theorem}{\ZaehlerbisEbene}{\glqq Satz\grqq} 
\DefThmUmgeb{Vermutung}{Theorem}{\ZaehlerbisEbene}
\DefThmUmgeb{Conjecture}{Theorem}{\ZaehlerbisEbene}
\DefThmUmgeb{Konvention}{Definition}{\ZaehlerbisEbene}
\DefThmUmgeb{Feststellung}{Theorem}{\ZaehlerbisEbene}
\DefUnterNumThmUmgeb{DefinitionZusatzNum}{Definition}{\ZaehlerbisEbene}{DefZN}{Definition}
\DefBspUmgeb{Beispiel}{Beispiel}{subsubsection}
\DefBspUmgeb{Example}{Example}{subsubsection}
\DefBspUmgeb{Frage}{Frage}{section}
\DefBspUmgeb{Question}{Question}{section}
\DefBspUmgeb{Aufgabe}{Aufgabe}{section}
\DefBspUmgeb{Ziel}{Ziel}{section}
\DefBemUmgeb{Bemerkung}
\DefBemUmgeb{Remark}
\DefSBemUmgeb{OffeneFrage}{Offene Frage}
\newcommand{\bdf}{\begin{Definition}}
\newcommand{\edf}{\end{Definition}}
\newcommand{\bvorsch}{\begin{Vorschrift}}
\newcommand{\evorsch}{\end{Vorschrift}}
\newcommand{\bconst}{\begin{Construction}}
\newcommand{\econst}{\end{Construction}}
\newcommand{\bthm}{\begin{Theorem}}
\newcommand{\ethm}{\end{Theorem}}
\newcommand{\bsatz}{\begin{Satz}}
\newcommand{\esatz}{\end{Satz}}
\newcommand{\bprop}{\begin{Proposition}}
\newcommand{\eprop}{\end{Proposition}}
\newcommand{\blem}{\begin{Lemma}}
\newcommand{\elem}{\end{Lemma}}
\newcommand{\bfolg}{\begin{Folgerung}}
\newcommand{\efolg}{\end{Folgerung}}
\newcommand{\bcorr}{\begin{Corollary}}
\newcommand{\ecorr}{\end{Corollary}}
\newcommand{\bfest}{\begin{Feststellung}}
\newcommand{\efest}{\end{Feststellung}}
\newcommand{\bbew}{\begin{Beweis}}
\newcommand{\ebew}{\end{Beweis}}
\newcommand{\bpf}{\begin{Proof}}
\newcommand{\epf}{\end{Proof}}
\newcommand{\bwnum}{\begin{WeiterNumerierteListe}}
\newcommand{\ewnum}{\end{WeiterNumerierteListe}}
\newcommand{\bdfzn}{\begin{DefinitionZusatzNum}}
\newcommand{\edfzn}{\end{DefinitionZusatzNum}}
\newcommand{\bbem}{\begin{Bemerkung}}
\newcommand{\ebem}{\end{Bemerkung}}
\newcommand{\brem}{\begin{Remark}}
\newcommand{\erem}{\end{Remark}}
\newcommand{\bnum}{\begin{NumerierteListe}}
\newcommand{\enum}{\end{NumerierteListe}}
\newcommand{\bunum}{\begin{UnnumerierteListe}}
\newcommand{\eunum}{\end{UnnumerierteListe}}
\newcommand{\bbsp}{\begin{Beispiel}}
\newcommand{\ebsp}{\end{Beispiel}}
\newcommand{\bex}{\begin{Example}}
\newcommand{\eex}{\end{Example}}
\newcommand{\bfrag}{\begin{Frage}}
\newcommand{\efrag}{\end{Frage}}
\newcommand{\bquest}{\begin{Question}}
\newcommand{\equest}{\end{Question}}
\newcommand{\baufg}{\begin{Aufgabe}}
\newcommand{\eaufg}{\end{Aufgabe}}
\newcommand{\bof}{\begin{OffeneFrage}}
\newcommand{\eof}{\end{OffeneFrage}}
\newcommand{\bverm}{\begin{Vermutung}}
\newcommand{\everm}{\end{Vermutung}}
\newcommand{\bconj}{\begin{Conjecture}}
\newcommand{\econj}{\end{Conjecture}}
\newcommand{\bkonv}{\begin{Konvention}}
\newcommand{\ekonv}{\end{Konvention}}
\newcommand{\bglklein}{\begin{StandardUnnumGleichungKlein}}
\newcommand{\eglklein}{\end{StandardUnnumGleichungKlein}}
\newcommand{\bgl}{\begin{StandardUnnumGleichung}}
\newcommand{\egl}{\end{StandardUnnumGleichung}}
\newcommand{\bglrtext}{\begin{XrelYZNumGleichung}}
\newcommand{\eglrtext}{\end{XrelYZNumGleichung}}

\newcommand{\berlgl}{\begin{StandardUnnumGleichung}}
\newcommand{\eerlgl}{\end{StandardUnnumGleichung}}
\newcommand{\beinrueck}{\begin{Einrueckungpur}} 
\newcommand{\eeinrueck}{\end{Einrueckungpur}}
\newcommand{\beinflist}{\begin{EinfachListe}} 
\newcommand{\eeinflist}{\end{EinfachListe}}
\newcommand{\beq}{\begin{equation}}
\newcommand{\eeq}{\end{equation}}
\newcommand{\bhin}{\begin{Hinrichtung}}
\newcommand{\ehin}{\end{Hinrichtung}}
\newcommand{\brueck}{\begin{Rueckrichtung}}
\newcommand{\erueck}{\end{Rueckrichtung}}
\newcommand{\bvl}{\begin{AutoLabelLaengenListe}{\ListNullAbstaende}}
\newcommand{\evl}{\end{AutoLabelLaengenListe}}
\newcommand{\df}[1]{{\bf #1}}

\newlength{\adressabstand}
\begin{document}
\title{On the Configuration Spaces of \\ Homogeneous Loop Quantum Cosmology \\ and Loop Quantum Gravity}
\author{Johannes Brunnemann$^{1,\ast}$ and Christian Fleischhack$^{1,2,}$\thanks{e-mail: 
            {\tt johannes.brunnemann@math.uni-hamburg.de}, {\tt christian.fleischhack@math.uni-hamburg.de}} \\   
        \\
        {\normalsize\em $^{1}$Department Mathematik}$^{\phantom{1}}$\\[\adressabstand]
        {\normalsize\em Universit\"at Hamburg}\\[\adressabstand]
        {\normalsize\em Bundesstra\ss e 55}\\[\adressabstand]
        {\normalsize\em 20146 Hamburg, Germany}
        \\[-25\adressabstand]      
        {\normalsize\em $^{2}$Max-Planck-Institut f\"ur Mathematik in den
                        Naturwissenschaften$^{\phantom{2}}$}\\[\adressabstand]
        {\normalsize\em Inselstra\ss e 22--26}\\[\adressabstand]
        {\normalsize\em 04103 Leipzig, Germany}
        \\[-25\adressabstand]}
\date{April 24, 2009}
\maketitle

\begin{abstract}
The set of homogeneous isotropic connections, as used in loop quantum 
cosmology, forms a line $l$ in the space of all connections $\A$.
This embedding, however, does not 
continuously extend to an embedding of the configuration space $\quer l$
of homogeneous isotropic loop quantum cosmology
into that of loop quantum gravity, $\Ab$.
This follows from the fact 
that the
parallel transports for general, non-straight paths 
in the base manifold 
do not depend almost periodically 
on $l$.
Analogous results are given for the anisotropic case.
\end{abstract}

%%%%%----------------------------------------------------------------%%%%%
%%%%%--------------- Section: Introduction        -------------------%%%%%
%%%%%----------------------------------------------------------------%%%%%

\section{Introduction}

In loop quantum cosmology (LQC), highly symmetric cosmological models
are quantized in analogy to loop quantum gravity (LQG).~\cite{d80,e90,u1}
One of the key features is the similar compactifications of their configuration spaces:
in LQG from $\A$ to $\Ab$ -- in LQC from $\R$ to $\rb$, the Bohr
compactification of $\R$. In both cases, 
``distributional'' objects are added to smooth ones:
added to smooth connections in LQG, to real numbers in LQC.
In fact, for homogeneous isotropic models (with $k = 0$), the classical configuration
space is spanned by $c \invzush$, where $c$ runs over $\R$ and $\invzush$ is
a fixed homogeneous and isotropic connection.
As in LQG, one does not consider these connections themselves,
but their parallel transports along certain edges.
Usually, only straight edges have been taken into account. 
For such edges $\gamma$ the parallel transports can be written down explicitly;
they equal
\bgl
h_{c \invzush} (\gamma) & = & \e^{- c \invzush(\dot \gamma) \laenge(\gamma)},
\egl
\noindent
where $\laenge(\gamma)$ denotes the length of $\gamma$.
In particular, they are periodic in $c$ (hence, almost periodic, as well)
and can be extended from $\R$ to $\rb$.
Another advantage of straight edges is that they separate the points in $\R$.
At the same time, however, the notion of
straightness requires a background metric.
Therefore, it seems appropriate to consider general, non-straight edges.
But, as we will prove in this note, these edges, in general, do not
lead to almost periodic parallel transports, whence they cannot be
extended continuously to $\rb$.
This will directly show that the configuration space of LQC
is not continuously embedded into that of LQG.
The importance of the question of the existence of this embedding was
highlighted in \cite{d83}.
General arguments extend our results to the anisotropic case.

\leerezeile

The paper is organized as follows: First we derive the 
differential equations for the par\-allel-transport matrix elements along
arbitrary curves. These are always second-order ODEs with a constant coefficient
in order 2, but generally non-constant and $c$-dependent coefficients
in orders 1 and 0. The coefficients encode, in particular, 
the underlying curve; they are
all constant w.r.t.\ the curve parameter iff this curve is a spiral arc.
Next, we introduce an ``invariant'' $\quasienergie$ for
the ODEs that (up to a constant) reduces to the standard energy invariant 
in the case of a harmonic oscillator. This generalized
energy is constant along the path up to ${\cal O}(\inv c)$. 
In the almost periodic case, $\quasienergie$ is fully invariant along the path.
Together with analyticity, this
imposes restrictions to the coefficients of the ODE
implying that the path is a straight line.
Finally, we find that the embedding $\R \nach \A$ is
not extendable to an embedding $\rb \nach \Ab$
since this required the almost periodicity of 
all parallel transports.
All the results are given for the homogeneous isotropic case with $k = 0$.
Since the almost periodicity on $\R^3$ implies the almost periodicity
on $\R$ being the diagonal in $\R^3$, the results extend to the 
general homogeneous case with $k = 0$. The paper concludes with an
appendix summarizing the main facts on almost periodic functions used here.

%%%%%----------------------------------------------------------------%%%%%
%%%%%--------------- Section: Introduction        -------------------%%%%%
%%%%%----------------------------------------------------------------%%%%%

\section{Preliminaries}
Let $P$ be a principal fibre bundle over a manifold
$M$ with compact structure Lie group $\LG$, and let $\invzush$ be a connection in $P$.
As we are aiming at homogeneous isotropic cosmology with $k = 0$, we assume the base
manifold to be contractible; hence $P$ is trivial.
Thus we may regard $\invzush$ as a $\Lieg$-valued $1$-form on $M$.
Moreover, again by triviality of the bundle, $c \invzush$ is a connection 
for every $c\in \R$.
Given an analytic edge $\gamma : I \nach M$ over an interval $I \teilmenge \R$ containing $0$
and using the standard trivialization of $P$,
we denote the parallel transport along $\gamma$ from $0$ to $t$
w.r.t.\ $c \invzush$
by $g_c(t) \in \LG$. The differential equation determining $g_c$ is
\bgl
\dot g_c(t) & = & - c \invzush (\dot\gamma(t)) \: g_c(t) \\
     g_c(0) & = & e_\LG.
\egl
The configuration space $\Ab$ of loop quantum gravity (or, more general, quantum geometry)
equals $\Hom(\Pf,\LG)$, where $\Pf$ 
is the groupoid of all analytic paths in $M$ (after modding out reparametrizations and immediate
retracings). It is given the initial topology induced by
the projections $\pi_\gamma : \Ab \nach \LG$ with $\gamma \in \Pf$ and 
$\pi_\gamma (\qa) := \qa(\gamma)$ assigning
to $\qa \in \Ab$ its parallel transport along $\gamma$. In an obvious manner, $\A$ is densely
embedded into $\Ab$. For further reading we refer to \cite{u1,diss}.
In loop quantum gravity, specifically, $M$ is a three-dimensional Cauchy slice and $\LG = SU(2)$.

%%%%%----------------------------------------------------------------%%%%%
%%%%%--------------- Section: Diff Eq SU(2)       -------------------%%%%%
%%%%%----------------------------------------------------------------%%%%%

\section{Differential Equations for $SU(2)$}
From now on, we consider the case of $M = \R^3$ and $\LG = SU(2)$ only. 
Moreover, $\invzush$ is a homogeneous and isotropic connection.
Let us write elements of $SU(2)$ as
\bgl
\bpm
\ph+ a & b \\ -\quer b & \quer a
\epm
\egl
\noindent
with $a,b \in \C$ fulfilling ${\betrag a}^2 + {\betrag b}^2 = 1$.
In the following, w.l.o.g., we assume that 
\bgl
\invzush & = & \tau_1 \dd x + \tau_2 \dd y + \tau_3 \dd z
\egl
\noindent
with
\bgl
\tau_1 = \bpm \ph+0 & -\I \\ -\I & \ph+0 \epm,
\breitrel{\breitrel{}}
\tau_2 = \bpm 0 & -1 \\ 1 & \ph+0 \epm,
\breitrel{\breitrel{}}
\tau_3 = \bpm -\I & 0 \\ \ph+0 & \I \epm.
\egl
\noindent
Consequently,
using $\gamma(t) = (x(t),y(t),z(t))$, 
we get
\bgl
\invzush(\dot\gamma(t)) 
  & = & \dot x  \tau_1 + \dot y \tau_2 + \dot z \tau_3
  \breitrel= \bpm -\I \dot z & - \I \dot x -\dot y \\ 
                  -\I \dot x +\dot y & \I \dot z \epm
  \breitrel{=} -\I \bpm n & \qm \\ \qqm & - n \epm \\
\egl
\noindent
with analytic
\bgl
   m & := & \dot x \vorz \I \dot y \\
   n & := & \dot z.
\egl
\noindent
In the following, we always assume that $\gamma$ is parametrized w.r.t.\
arc length, i.\,e.,
\bgl
{\betrag \qqm}^2 + n^2 \breitrel= \norm{\dot\gamma}^2 \breitrel\ident 1.
\egl
The equation for the parallel transport reads now for each $c$
\begin{eqnarray}
\label{eq:dgl_pt}
\bpm
\ph+ \dot a & \dot b \\ -\dot{\quer b} & \dot{\quer a}
\epm
& = & 
\I c
\bpm n & \ph-\qm \\ \qqm & - n \epm
\bpm
\ph+ a & b \\ -\quer b & \quer a
\epm
\end{eqnarray}
\noindent
or
\begin{eqnarray}
\label{eq:dgl_apunkt}
\dot a & = & \I c (na - \qm \quer b) \\
\dot b & = & \I c (nb + \qm \quer a) 
\end{eqnarray}
\noindent
with the initial conditions
\begin{eqnarray}
a(0) & = & 1 \nonumber \\
b(0) & = & 0. \nonumber
\end{eqnarray}

From that we get for $m \neq 0$
\begin{eqnarray}
\ddot a & = & \I c (\dot n a + n \dot a 
                   - \dot{\qm} \quer b - \qm \dot{\quer b}) \nonumber\\
        & = & \I c (\dot n a + n \dot a 
                   - \dot{\qm} \quer b) - c^2 \qm (\qqm a + n\quer b) \nonumber\\
        & = & \textstyle \I c (\dot n a + n \dot a)  
                   - c^2 \qm \qqm a
                   - (\I c \dot{\qm} + c^2 \qm n)\inv{\qm} \bigr(na - \inv{\I c} \dot a\bigl)\nonumber\\
\label{eq:dgl_a}
        & = & \I c (\dot n - Mn) a 
                   - c^2 a 
                   + M \dot a \hspace*{5ex} \erl{since $\qm \qqm + n n = 1$}
\label{eq:dgl2_a}
\end{eqnarray}
\noindent
with
\bgl
M & := & \frac{\dot \qm}{\qm} \:.
\egl

Analogously, we get 
\begin{eqnarray}
\ddot b 
  & = & \I c (\dot n b + n \dot b 
               + \dot{\qm} \quer a + \qm \dot{\quer a}) \nonumber\\
  & = & \I c (\dot n b + n \dot b 
               + \dot{\qm} \quer a)
         - c^2 \qm (\qqm b - n \quer a) \nonumber\\
  & = & \textstyle \I c (\dot n b + n \dot b) 
         - c^2 \qm \qqm b 
           +( \I c \dot{\qm} + c^2 \qm  n) 
              \inv{\qm}\bigl(\inv{\I c} \dot b - nb\bigr) \nonumber\\
  & = & \I c (\dot n - M n) b
         - c^2 b + M \dot b.
\label{eq:dgl2_b}
\end{eqnarray}
\noindent

%%%%%----------------------------------------------------------------%%%%%
%%%%%--------------- Section: Further Transformation      -----------%%%%%
%%%%%----------------------------------------------------------------%%%%%

\section{An Energy-like Invariant}
The central object we will exploit to prove the non-almost periodicity result
is given by
\bdf
\bgl
\quasienergie_a(c,t) 
 & := & \bigl(\qqm a^2 + 2 na\quer b - \qm \quer b^2\bigr)\big|_0^t
\egl
\edf\noindent
It
generalizes the energy conservation functional known from the special case of the
differential equation $\ddot x + \omega^2 x = 0$ describing an harmonic oscillator.
In fact, again assuming $m \neq 0$, we have by \eqref{eq:dgl_apunkt}
and $\qm \qqm + n n = 1$
\bgl
\hspace*{-3em}
\quasienergie_a(c,t) 
 \breitrel= \bigl(\qqm a^2 + 2 na\quer b - \qm \quer b^2\bigr)\big|_0^t
 & = & \inv\qm \Bigl(-(na - \qm \quer b)^2 + a^2\Bigr)\Big|_0^t  %\\
 \breitrel= \inv\qm \Bigl(\frac{\dot a^2}{c^2} + a^2\Bigr)\Big|_0^t\:.
\hspace*{-3em}
\egl

In order to eliminate the disturbing $\inv\qm$-term and to remove the first-order term
in our differential equations \eqref{eq:dgl2_a} and \eqref{eq:dgl2_b}, we decompose $a$ as usual 
into%
\footnote{The choice of the square-root leaf is made continuously w.r.t.\ $t$.
Moreover, throughout this section we continue to assume $m \neq 0$.}
\newcommand{\faf}{\alpha}
\bgl
a & = & \sqrt{\qm} \: \faf.
\egl\noindent
This implies
\bglklein
\sqrt{\qm} \: a & = & \qm \faf, \\
\sqrt{\qm} \: \dot a & = & \inv2 \dot{\qm} \faf + \qm \dot\faf, \\
\sqrt{\qm}^3 \:\ddot a 
   & = & \inv2 \qm \ddot {\qm} \faf - \inv4 \dot{\qm}^2 \faf
            +  \qm \dot {\qm} \dot\faf + \qm ^2 \ddot\faf.
\eglklein
Now, the differential equation for $a$ transforms into
\bglklein
0  
  & = & \sqrt{\qm}^3 \: \bigl(\ddot a - M \dot a + c^2 a - \I c (\dot n - Mn) a\bigr) \\
  & = & \inv2 \dot M \qm^2  \faf - \frac14 \dot{\qm}^2 \faf
           + \qm ^2 \ddot\faf
       + c^2 \qm ^2\faf
                - \I c (\dot n - Mn) \qm ^2 \faf \\
\eglklein\noindent
and
\bglklein
\ddot\faf + c^2 \faf
  & = & \bigl(\frac14 M^2 - \inv2 \dot M 
                + \I c (\dot n - Mn) \bigr)  \faf . \\
\eglklein\noindent

With $\ffa := \frac14 M^2 - \inv2 \dot M$ and $\ffb := \I (\dot n - Mn)$,
we have
\bgl
\frac\dd{\dd t}\Bigl(\frac{\dot \faf^2}{c^2} + \faf^2\Bigr)
  & = & \frac2{c^2} \bigl(\ddot\faf + c^2 \faf \bigr) \dot\faf 
  \breitrel= \frac1{c^2} \bigl(\ffa + c \ffb \bigr)  \frac\dd{\dd t} \faf^2,
\egl\noindent
hence
\bgl[2ex]
\quasienergie_\faf(c,t) \breitrel{:=} \Bigl(\frac{\dot \faf^2}{c^2} + \faf^2\Bigr) \Big|_0^t
  & = & \frac1{c^2} \int_0^t\bigl(\ffa + c \ffb \bigr)  \frac\dd{\dd \tau} \faf^2 \: \dd \tau \s
  & = & - \frac1{c^2} \int_0^t\bigl(\dot\ffa + c \dot\ffb \bigr)  \faf^2 \: \dd \tau 
        + \frac1{c^2} \bigl(\ffa + c \ffb \bigr)  \faf^2  \Big|_0^t. \\
\egl\noindent
For each $t$, the functions $\rho$ and $\sigma$,
as well as their derivatives, are bounded on $[0,t]$. 
Since $\qm$ is independent of $c$ and does, moreover, nowhere
vanish, we have $\sup_{[0,t]}\betrag{\faf} \leq C$ 
with a $c$-independent constant $C$. (Recall $\supnorm a = 1$.)
This immediately implies that 
\bgl
\lim_{c \gegen \infty} \quasienergie_\faf(c,t) 
 & = & 0
\egl\noindent
for each parameter value $t$.
Rewriting $\dot\faf$ in terms of $a$ and $\dot a$, we get
\bglklein
\dot\faf^2 
  & = &  \inv{\qm^2}\bigl(\sqrt \qm \: \dot a - \inv2 \dot\qm \faf\bigr)^2 
  \breitrel= \inv{\qm}\bigl(\dot a^2 - M a \dot a + \inv4 M^2 a^2\bigr),
\eglklein\noindent
and we have
\bgl
\quasienergie_a(c,t) 
 & \ident &  
               \inv\qm \Bigl(\frac{\dot a^2}{c^2} + a^2\Bigr) \Big|_0^t 
 \breitrel= \quasienergie_\faf(c,t) + \frac{Ma}{\qm c^2} \Bigl(\dot a - \inv4 M a\Bigr) \Big|_0^t .
\egl\noindent
Due to 
\bgl
\supnorm{\dot a} & \leq & c (\supnorm n + \supnorm \qm)
\egl\noindent
we have
\blem
\bgl
\lim_{c \gegen \infty} \quasienergie_a(c,t) 
 & = & 0
\egl\noindent
\elem

\bcorr
\label{corr:konst_energie}
If, for some parameter value $t$, the functions $a(t)$ and $b(t)$ are 
almost periodic w.r.t.\ $c$, then $\quasienergie_a(c,t) = 0$ for all $c$.
\ecorr
For the definition of almost periodicity, see Appendix \ref{app:alm-per}.

\bpf
If $a(t)$ and $b(t)$ are almost periodic, 
then
$\quasienergie_a(\cdot,t)$ is almost periodic as well.
Since (cf.\ Lemma \ref{lem:converg_apf})
any almost periodic function is already constant, provided
it converges while the argument
is approaching $\infty$,
 we get the proof.
\qed
\epf
Of course, the preceding corollary remains correct if we replace 
$a$ by $b$.

From now on, we call $t$ sloppily almost periodic iff $a(t)$ and $b(t)$
depend almost periodically on $c$. This yields

\blem
\label{lem:ident_konst_energie}
If the set of almost periodic parameter values contains
an accumulation point, then $\quasienergie_a \ident 0$.
\elem
\bpf
For general reasons \cite{EMS1}, %I.2.6 auf Seite 17
$\quasienergie_a$ depends locally (real) analytically on ($c$ and) $t$.
Now, for each $c$, the function $t \auf \quasienergie_a(c,t)$ has an accumulation
point of zeros by the preceding corollary.
Consequently, this function vanishes
identically for each $c$.
\qed
\epf

Next, we would like to investigate the derivative of $\quasienergie_a$ w.r.t.\ $t$.
For that, we consider 
\begin{eqnarray}
\label{eq:allgF}
 F & := & f_2 a^2 + f_1 a \quer b + f_0 \quer b^2,
\end{eqnarray}\noindent
where $f_0, f_1, f_2$ may depend on $t$, but not on $c$.
We have 
\begin{eqnarray}
 \dot F 
    & = & \bigl(\dot f_2 + \I c (2 f_2 n - f_1 \qqm)\bigr) \, a^2 
	       + \bigl(\dot f_1 - 2 \I c (f_2 \qm + f_0 \qqm)\bigr) \, a \quer b 
               + \bigl(\dot f_0 - \I c (2 f_0 n + f_1 \qm)\bigr) \, \quer b^2.
\nonumber
\end{eqnarray}\noindent
If $a$ and $b$ are almost periodic, it follows that 
\bgl
\dot f_2 a^2 + \dot f_1 a \quer b + \dot f_0 \quer b^2
\egl\noindent
is almost periodic, implying for almost periodic $\dot F$ that
\bgl
\I c \bigl((2 f_2 n - f_1 \qqm) \: a^2 
	       - 2 (f_2 \qm + f_0 \qqm) \: a \quer b 
               - (2 f_0 n + f_1 \qm) \: \quer b^2\bigr) 
\egl\noindent
is almost periodic as well. Consequently
(cf.\ Lemma \ref{lem:f_cf_0}),
the almost periodic function
\bgl
(2 f_2 n - f_1 \qqm) \: a^2 
	       - 2 (f_2 \qm + f_0 \qqm) \: a \quer b 
               - (2 f_0 n + f_1 \qm) \: \quer b^2 
\egl\noindent
is even identically zero. This, finally, implies
that 
\begin{eqnarray}
\label{eq:dotF}
 \dot F & = & \dot f_2 a^2 + \dot f_1 a \quer b + \dot f_0 \quer b^2,
\end{eqnarray}\noindent
for almost periodic $a$, $b$ and $\dot F$.

\blem
\label{lem:abl-koeff}
If the set of almost periodic parameter values contains
an accumulation point, then
we have
\bgl
 \dot\qqm a^2 + 2 \dot n a\quer b - \dot\qm \quer b^2 & = & 0.
\egl
\elem
\bpf

Setting
\bgl
f_2 & := & \qqm \\
f_1 & := & 2n \\
f_0 & := & -\qm
\egl
into \eqref{eq:allgF},
we get $\quasienergie_a(c,t) = F(c,t) - F(c,0)$.  
The assertion now follows from \eqref{eq:dotF} 
and Lemma \ref{lem:ident_konst_energie}.
\qed
\epf
We note that, by induction, one gets
$\qqm^{(r)} \, a^2 + 2 n^{(r)} \, a\quer b - \qm^{(r)} \, \quer b^2 = 0$
for all non-zero $r\in\N$.
Moreover, the equation for $r = 1$ 
stated in the lemma above could also have been obtained directly,
as for that choice of $f_0, f_1, f_2$ the $\I c$-term already vanishes without the almost-periodicity assumption.

\bcorr
\label{corr:acc+ap=line}
Any curve having an
accumulation point of almost periodic
parameter values is a straight line. 
\ecorr

\bpf
Assume there is an almost periodic 
$t$ with $\betrag{\dot\qm(t)} \neq 0$. 
As $\dot\qm$ is continuous, we have $\dot\qm \neq 0$ on some open $U$ containing $t$.
Moreover, Lemma \ref{lem:ident_konst_energie} and Lemma \ref{lem:abl-koeff} show
\bgl
    \qqm a^2 + 2      na\quer b -     \qm \quer b^2 & = & \qqm(0) \\
\dot\qqm a^2 + 2 \dot na\quer b - \dot\qm \quer b^2 & = & 0. \\
\egl
Next, a short calculation shows that for either sign
\bgl
\Biggl(2 \frac{\dot n^2}{\dot\qqm{}^2} 
      - \frac{\qm}{\qqm} + \frac{\dot\qm}{\dot\qqm} 
      - 2 \frac{n \dot n}{\qqm \dot\qqm}
      \pm 2 \Bigl(\frac{n}{\qqm} - \frac{\dot n}{\dot\qqm} \Bigr)
            \sqrt{\frac{\dot n^2}{\dot\qqm{}^2} + \frac{\dot\qm}{\dot\qqm}} \:
      \Biggr) \: \quer b^2 
    & = & \frac{\qqm(0)}{\qqm},
\egl
i.e., $\quer b^2$ does not depend on $c$ all over $U$---and neither
does $b$. (The large bracket term cannot be zero, because this contradicts
the assumption that $m \neq 0$ everywhere.) The same is true for $a$.
Consequently, $\del_c a$ and $\del_c b$ vanish on $U$.
Using now the original differential equations,
we get on $U$
\bgl
0 & = & \del_t \del_c a 
 \breitrel\ident \del_c \dot a 
 \breitrel= \I (na - \qm \quer b) +
             \I c (n \del_c a - \qm \del_c \quer b) 
 \breitrel= \I (na - \qm \quer b) \\
0 & = & \del_t \del_c b 
 \breitrel\ident \del_c \dot b 
 \breitrel= \I (nb + \qm \quer a) + \I c (n \del_c b + \qm \del_c \quer a) 
 \breitrel= \I (nb + \qm \quer a).
\egl
This implies
\bgl
a & = & (n^2 + \qqm\qm) a 
  \breitrel= n\qm\quer b + \qqm\qm a \breitrel= 0
\egl
and thus $b = 0$ on $U$. This, of course, is a contradiction.
Consequently, $\dot m \ident 0$, whence also $\dot n \ident 0$, 
and $\gamma$ is a line.
\qed
\epf
%%%%%----------------------------------------------------------------%%%%%
%%%%%--------------- Section: Special Cases       -------------------%%%%%
%%%%%----------------------------------------------------------------%%%%%

\section{Non-Almost Periodicity}
\bthm
\label{thm:isotrop_nichtfp}
Let $\gamma$ be an analytic curve that is not part of a 
straight line.

Then there is a $T>0$, such that the parallel transport $g_c(t)$
along $\gamma$ is not almost periodic w.r.t.\ $c$ for any $0 < t < T$.
\ethm
\bpf
The case $m(0) \neq 0$ is proven in Corollary \ref{corr:acc+ap=line}.
If $m(0) = 0$, choose some $h \in SU(2)$, such that none
of its matrix elements vanishes. Conjugating \eqref{eq:dgl_pt} by $h$,
we get a new differential equation of the same type; however, $m$
is replaced by some function of $m$, $n$ and $h$, not vanishing at $0$.
Since the property of being almost periodic does not change under
conjugation with a fixed element, we get the proof.
\qed 
\epf
\brem
At present, only accumulation points of 
almost periodic parameter values can be excluded for nonlinear curves.
It is still unclear whether there might exist isolated 
almost periodic parameter values. Nevertheless we do not expect this 
to be case; for paths that are spiral arcs, this will indeed be proven in the next
section. If we were right in general, 
the non-almost periodicity in the theorem above
would be given for all $0 < t \leq \laenge(\gamma)$.
\erem

%%%%%----------------------------------------------------------------%%%%%
%%%%%--------------- Section: Special Cases       -------------------%%%%%
%%%%%----------------------------------------------------------------%%%%%

\section{Special Cases: Spiral Arcs}
\label{sect:special_cases}
The differential equation \eqref{eq:dgl_a}
for $a$ has, in general, $t$-dependent
coefficients. Nevertheless, there are special cases where the
coefficients are independent of $t$: lines and circles, or more generally,
spirals. They, on the other hand, will turn out to be the only possible examples
for $m \neq 0$. 
Indeed, constancy of coefficients implies
$M \ident \qquer\kappa$ for some $\kappa \in \C$.
Therefore $\dot m = \kappa m$ or $m = C \e^{\kappa t}$ 
for some non-zero $C \in \C$. There remain two cases:

\subsubsection*{Case 1: \: $\kappa = 0$}

Then $m$ is constant, whence $n^2$ is constant implying that $n$
is constant as well. Altogether $\dot\gamma$ is constant,
whence $\gamma$ is a line.

\subsubsection*{Case 2: \: $\kappa \neq 0$}

Then $\dot n - Mn = \dot n - \kappa n$ is constant, say $\Lambda$. Thus,
$n = -\frac \Lambda \kappa + D \e^{\kappa t}$ for some $D \in \C$. Since $n$ and its derivatives
$\dot n(0) = \kappa D$ and $\ddot n(0) = \kappa^2 D$ are real, we deduce
$D$ is real.

\leerezeile

\bvl{}
\iitem[{\em Subcase 2a:}]
$D \neq 0$.\leerezeile

Then $\kappa$ is real (since $\kappa D$ is real) 
and therefore $\Lambda$ as well (recall that $D$ and $n$ are real).
We get for all parameter values $t$
\bgl[2ex]
1 & = & \betrag m ^2 + n^2 
  \breitrel= \betrag{C}^2 \e^{2\kappa t} + \Bigl(-\frac \Lambda \kappa + D \e^{\kappa t} \Bigr)^2 \s
  & = & (\betrag{C}^2 + D^2) \e^{2\kappa t} - 2 D \frac \Lambda \kappa \e^{\kappa t} + \frac{\Lambda^2}{\kappa^2}. 
\egl
This requires $\betrag C = D = 0$, implying $n = \pm 1$ and $m \ident 0$,
in contradiction to the assumption above.

\leerezeile
\iitem[{\em Subcase 2b:}]
$D = 0$.\leerezeile

Then $n$ is constant. From
\bgl
1 & = & \betrag m ^2 + n^2 
  \breitrel= \betrag{C}^2 \e^{2 (\re \kappa) t} + \frac{\Lambda^2}{\kappa^2} \\
\egl
we deduce that $\re \kappa = 0$. Thus $\kappa = \I \lambda$ for some 
non-zero $\lambda \in \R$,
and $m = C \e^{\I\lambda t}$ with $\betrag C \leq 1$. Therefore
\bgl
x - \I y & = & \frac{C}{\I\lambda} \e^{\I\lambda t} + E
\egl
with some $E \in \C$.
$n$ is constant with 
$n^2 = 1 - \betrag C ^2$.
Geometrically, $\gamma$ is (part of) a spiral, constantly
moving in $z$-direction which projects down to (part of) a circle 
in the $x$-$y$-plane.

\evl

\leerezeile

Altogether, we see that the only possible cases for constant coefficients
are edges whose tangents are constant in $z$-direction and
project down to a circle or a line in the $x$-$y$-plane.
In these cases, the differential equation looks like
\begin{eqnarray}
\label{eq:einfach_a}
\ddot a + \I \lambda \dot a + (c^2 - c \varkappa) a & = & 0
\end{eqnarray}
\noindent
with $\lambda, \varkappa \in \R$, subject to the condition
$\betrag\varkappa \leq \betrag\lambda$. 
The case $\lambda = 0$ (and, consequently, $\varkappa = 0$, as well)
corresponds to the case of a line.
In general, the case $\varkappa = 0$ represents
edges lying in a plane perpendicular to the $z$-axis, unless $\lambda = 0$.
The equation for $b$ reduces in the constant-coefficient case to
\begin{eqnarray}
\ddot b + \I \lambda \dot b + (c^2 - c \varkappa) b & = & 0.
\end{eqnarray}

%%%%%----------------------------------------------------------------%%%%%
%%%%%--------------- Section: General Solution    -------------------%%%%%
%%%%%----------------------------------------------------------------%%%%%

The general solution for Eq.\ \eqref{eq:einfach_a} is
\begin{eqnarray}
\nonumber
a(t) 
  & = & \e^{-\I\frac\lambda2 t}
          \Bigl(a(0) \cos \sq t + 
	       \inv{\sq}\bigl(\dot a(0) + \I \frac\lambda2 a(0)\bigr) \sin \sq t \Bigr)
\end{eqnarray}
with
\begin{eqnarray}
\nonumber
\sq & = & \sqausgeschrieben \breitrel= \sqandersausgeschrieben.
\end{eqnarray}
Hence, we get for $\sq \neq 0$
with
\bgl
\dot a(0) & = & \I c n(0) \\
\dot b(0) & = & \I c \qm(0)
\egl\noindent
the solutions
\begin{eqnarray}
\nonumber
a(t) 
  & = & \e^{-\I\frac\lambda2 t}
          \Bigl(\cos \sq t + 
	       \frac\I{\sq}\bigl(c n(0) + \frac\lambda2 \bigr) \sin \sq t \Bigr) \\
\nonumber
b(t) 
  & = & \e^{-\I\frac\lambda2 t} \: \frac\I{\sq} c \qm(0) \sin \sq t.
\end{eqnarray}

In the case that $\gamma$ is a line, $\lambda$ and $\varkappa$ are zero.
With $\sq = \betrag c$, we get
\blem
\label{lem:line_ap}If $\gamma$ is a line, then
\begin{eqnarray}
\nonumber
a(t) 
  & = & \cos c t + \I n(0) \sin c t \\
\nonumber
b(t) 
  & = & \I \qm(0)  \sin c t.
\end{eqnarray}
In particular, these functions are periodic. 
\elem

%%%%%----------------------------------------------------------------%%%%%
%%%%%--------------- Section: Criterion for Almost Periodicity ------%%%%%
%%%%%----------------------------------------------------------------%%%%%

In the remaining part of this section we are going to show that the
lines are the only type of 
spiral arcs having almost periodic parallel transports.
Others spiral arcs cannot even have isolated almost periodic parallel transports
answering the open question in the remark above, at least for this type of paths.

We start with a criterion for non-almost periodicity.
\blem
\label{lem:crit_almper}
Let $f : \R \nach \R$ be a continuous 
function. Assume there is an interval $I$
in $\R$ and some $r,\varepsilon > 0$, 
such that 
\begin{eqnarray}
\label{eq:absch}
\sup_J \betrag{f} - \sup_{I} \betrag{f} & \geq & \varepsilon 
\end{eqnarray}
for all intervals 
$J$ with $\inf J > r$ and $\betrag J = \betrag I$.
Then $f$ is not almost periodic.
\elem

\bpf
Assume $f$ is almost periodic. 
Then there is some $L>0$ (w.l.o.g., we may choose $L > r + \betrag{\inf I}$), 
such that
there is some $\omega \in [L,2L]$
with $\betrag{f(c) - f(c-\omega)} < \varepsilon$ for all $c\in\R$.
Define $J := I + \omega$. By assumption, 
$\inf J = \inf I +\omega \geq -\betrag{\inf I} + L > r$.
Choose now $C\in J$ with $\betrag{f(C)} = \sup_J \betrag f$.
Since $C-\omega \in I$, we have

\bgl
\betrag{f(C) - f(C-\omega)} 
  \breitrel\geq \betrag{f(C)} - \betrag{f(C-\omega)}
  \breitrel\geq \sup_J \betrag f - \sup_I \betrag f 
  \breitrel\geq \varepsilon
\egl
by assumption \eqref{eq:absch}. Contradiction.
\qed 
\epf

\bprop
Let $t,\lambda,\varkappa \in \R$ with  
$\betrag\varkappa\leq\betrag\lambda$.
Then the function $\fff : \R \nach \R$, given by  
\bgl
\fff(c) 
  & := & 
         \frac c{\sqausgeschrieben}  \sin \sqausgeschrieben \: t \\
\egl
is almost periodic iff $\lambda = 0$ or $t = 0$.
\eprop
Here, in the case of $\betrag\varkappa = \betrag\lambda$,
we set $\fff(\frac\varkappa2) := ct$ to get a continuous function.

\bpf
\bunum
\item
W.l.o.g., we may assume that $\varkappa \leq 0$ and $\lambda,t\geq 0$.
Since the cases $\lambda = 0$ 
(see Lemma \ref{lem:line_ap})
and $t = 0$ are 
obvious,
we even may assume $\lambda,t > 0$.
\item
Possibly ignoring $c = 0$, the zeros of $\fff$ are given by
\begin{eqnarray}
\label{eq:zeros}
c_{\pm k} & := & \frac\varkappa2 \pm \sqrt{\frac{\pi^2 k^2}{t^2} - \frac{\lambda^2-\varkappa^2}4}
\end{eqnarray}
for positive integers $k\geq k_0$ with
\bgl
k_0 & := & \Bigogkl{\frac{t}{2\pi} \sqrt{\lambda^2-\varkappa^2}}.
\egl
We have $c_k < c_{k+1}$ for all $k$, and one easily checks that 
\begin{eqnarray}
\nonumber
\lim_{k\gegen\infty} (c_{k+1} - c_k) & = &  \frac{\pi}{t}.
\end{eqnarray}
\item
Observe that 
the function
\bgl
\ffh(c) & := & \frac c{\sqausgeschrieben} 
\egl
is strictly monotonously 
increasing on $\R_+$ with $\lim_{c\gegen\infty} \ffh(c) = 1$.
Defining $I_k := [c_k,c_{k+1}]$ for $k \geq k_0$, we have
from the sine properties
\bgl
 \sup_{I_k} \betrag{\fff}
  \breitrel\geq \betrag{\fff(c_{k+\inv2})}
  \breitrel= \ffh(c_{k+\inv2})
  \breitrel\geq \inf_{I_k} \ffh 
  \breitrel= \ffh(c_{k})
\egl
for $k \geq k_0$, where we have naturally extended \eqref{eq:zeros}
to half-integer indices. 
Moreover, for each interval $I$ in $\R_+$
\bgl
\sup_{I} \betrag{\fff}
  \breitrel\leq \sup_{I} \ffh
  \breitrel= \ffh(\sup I).
\egl
\item
Now, let $I := [c_{k_0},c_{k_1}]$ 
with $k_1 \in \N_+$ and $\betrag I \geq 3 \frac{\pi}t$.
Next, choose some $k_\varrho\in\N_+$ with $c_{k+1} - c_k < \frac32 \frac{\pi}t$ for all 
integer $k \geq k_\varrho$. Choose, finally, $r > \max\{c_{k_1},c_{k_\varrho}\}$
and let
$\varepsilon := \ffh(r) - \ffh(c_{k_1}) > 0$.

Then each interval $J = [c,c+\betrag I]$ with $c > r$ contains
at least one complete interval $I_k = [c_k,c_{k+1}]$
with $k\geq k_\varrho$.
In fact, 
the half-interval $[c_{k_\varrho},\infty) \echteobermenge [r,\infty)$ 
containing $J$ is covered by the intervals
$I_i$, $i \geq k_\varrho$, each having length less than $\frac32 \frac{\pi}t$. 
On the other hand, the length of $J$ 
equals the length of $I$ being at
least $3\frac{\pi}t$.

This implies (recall $c_k > r$)
\bgl
\sup_J \betrag{\fff} - \sup_I \betrag{\fff} 
  & \geq & \sup_{I_k} \betrag{\fff} - \ffh(c_{k_1})
  \breitrel\geq \ffh(c_{k}) - \ffh(c_{k_1}) \\
  & > & \ffh(r) - \ffh(c_{k_1}) 
  \breitrel= \varepsilon.
\egl
Lemma \ref{lem:crit_almper} gives the proof.
\qed
\eunum
\epf

\bcorr
The parallel transport along a path $\gamma$ being (part of) 
a line or a spiral arc 
is almost periodic as a 
function of $\R$ into $\LG = SU(2)$ iff $\gamma$ is (part of) a line.
\ecorr

%%%%%----------------------------------------------------------------%%%%%
%%%%%--------------- Section: LQC in LQG          -------------------%%%%%
%%%%%----------------------------------------------------------------%%%%%

\section{Relating the Configuration Spaces of LQC and LQG}
\label{sect:relate_configs}
Now we are prepared to answer the
question, whether the configuration space of loop quantum cosmology
is continuously embedded into the configuration space of 
loop quantum gravity.
More precisely, can the
canonical embedding
\bgl
\fktdefabgesetzt{\iota}{\R}{\A}{c}{c \invzush}
\egl\noindent
with a fixed homogeneous isotropic $\invzush$ be continuously 
extended to an embedding 
\bgl
\fktabgesetzt{\quer\iota}{\rb}{\Ab\:?}
\egl\noindent
If this were the case, the mapping 
\bgl
\fktabgesetzt{\iota_{\Ab} \circ \iota}{\rb \obermenge \R}{\A \breitrel\nach \Ab}
\egl\noindent
had to be continuous (with $\R$ given the topology inherited from $\rb$
and $\iota_{\Ab}$ being the standard embedding of $\A$ into $\Ab$).
From basic theorems on projective limits \cite{diss}
it follows that this is equivalent
to the continuity of 
\bgl
\fktabgesetzt{\pi_e \circ \iota_{\Ab} \circ \iota}{\rb \obermenge \R}{\A \breitrel\nach \Ab \breitrel\nach \LG}
\egl\noindent
for all edges $e$ in $M$. 
A map from $\R$ to $\LG$, on the other hand,
is continuous w.r.t.\ the Bohr topology
iff it is almost periodic \cite{RudinFour}.%
\footnote{In \cite{RudinFour} the case of complex-valued functions has been considered.
The general statement for $\LG$-valued functions 
follows from Appendix \ref{app:alm-per} together with the fact that a map 
to $\LG$ is continuous iff all of its corresponding 
matrix-element functions are so. Recall that any compact Lie group is 
(embedded into) a matrix group.}
However, since, as seen above, there are edges $e$ that 
yield non-almost periodic functions 
\bgl
\fktdefabgesetzt{\pi_e \circ \iota_{\Ab} \circ \iota}{\R}{\LG,}{c}{h_e(c \invzush)}
\egl\noindent
we get a contradiction.

Consequently, the embedding $\iota$ of $\R$ into $\A$ cannot be
continuously extended to an embedding of $\rb$ into $\Ab$,
i.e., of the configuration space of loop quantum cosmology into that
of loop quantum gravity.

%%%%%----------------------------------------------------------------%%%%%
%%%%%--------------- Section: Anisotropic Case    -------------------%%%%%
%%%%%----------------------------------------------------------------%%%%%

\section{Anisotropic Case}

In the anisotropic case one does not consider the connections $c \invzush$
with $\invzush$ being homogeneous isotropic, but the connections

\bgl
A_{\vc} & = & c_1 \tau_1 \dd x + c_2 \tau_2 \dd y + c_3 \tau_3 \dd z
\egl
\noindent
with $\vc = (c_1, c_2, c_3) \in \R^3$. Of course, 
$A_{\vc}$ is isotropic iff all components of $\vc$ coincide.
This observation allows us to extend the results above to the
anisotropic case thanks to
\blem
Let $f : \R^3 \nach \C$ be an almost periodic function. 

Then
$f^\diag : \R \nach \C$ with $f^\diag(x) := (x,x,x)$ is almost periodic 
as well.
\elem
\bpf
$f$ is almost periodic iff \cite{RudinFour} it is 
a uniform limit $\lim_{n \gegen \infty} f_n$ of trigonometric polynomials.
Since each $(f_n)^\diag$ is, of course, a trigonometric polynomial again and since
\bgl
\sup_\R \betrag{f^\diag - (f_n)^\diag}
 & \ident & \sup_{(x,x,x) \in \R^3} \betrag{f - f_n}
 \breitrel\leq \sup_{\R^3} \betrag{f - f_n}
 \breitrel\gegen  0,
\egl
$f^\diag$ is a uniform limit of trigonometric polynomials,
hence almost periodic as well.
\qed
\epf
Now we have from Theorem \ref{thm:isotrop_nichtfp}

\bthm
Let $\gamma$ be an analytic curve that is not part of a 
straight line. 

Then there is a $T>0$, such that  
\bgl
g_\vc(t) & := & \text{parallel transport w.r.t.\ $A_\vc$ along $\gamma$ from $\gamma(0)$ to $\gamma(t)$}
\egl
is not almost periodic w.r.t.\ $\vc$ for any $0 < t < T$.
\ethm
Completely analogously to Section \ref{sect:relate_configs}, we 
see that the embedding 
\bgl
\fktdefabgesetzt{\iota_\aniso}{\R^3}{\A}{\vc}{A_\vc}
\egl\noindent
cannot be continuously 
extended to an embedding 
\bgl
\fktabgesetzt{\quer\iota_\aniso}{\rb^3}{\Ab\:.}
\egl\noindent

\brem
Both in the isotropic and in the anisotropic case,
we expect all main results 
to remain true 
if considering non-zero $k$, i.e.\
spherical or hyperbolic universes. Of course, then, straight lines
have to be replaced by geodesics.
\erem

\section{Notes Added in Proof}
As communicated to us by Martin Bojowald, Tim Koslowski
(for details, see \cite{d84})
has given power series expansions for the solutions in the case
of planar edges ($n = 0$).
These indicate that each of these parallel transports can be written as a 
sum of a function being (almost) periodic in $c$ 
and a function vanishing for large $\betrag c$. This is confirmed by our
explicit solutions in the constant-coefficient case.

\section*{Acknowledgements}

The authors are very grateful to Jonathan Engle for raising the
question on the embeddability of LQC into LQG during his visits in 
Hamburg. Moreover, the authors would like to thank Tim Koslowski for 
fruitful discussions.
The work has been supported by the Emmy-Noether-Programm 
(grant FL~622/1-1) of the Deutsche Forschungsgemeinschaft.

\anhangengl

\section{Basics on Almost Periodic Functions}
\label{app:alm-per}

\bdf
A function $f : \R \nach \C$ is called \df{almost periodic} 
iff it is continuous
and for any
$\varepsilon > 0$ there is an $L > 0$, such that each open interval in $\R$ 
of length $L$ contains a $\xi$, such that
$\betrag{f(x + \xi) - f(x)} \leq \varepsilon$ for all $x \in \R$. \cite{Bohr}
\edf
Constant functions are almost periodic; sum and products of almost
periodic functions are almost periodic as well.
Almost periodic functions are always bounded.
\blem
\label{lem:converg_apf}
Let $f : \R \nach \C$ be 
an
almost periodic function,
such that $\lim_{c\gegen\infty} f(c)$ exists and equals 
$f_\infty\in\C$. Then $f$ is constantly equal $f_\infty$.
\elem
\bpf
Let $\varepsilon > 0$.

By assumption, 
there is an $L>0 $, 
such that $\betrag{f(c) - f_\infty} < \inv2\varepsilon$ for all
$c > L$.
Assume there is some $C \in \R$, such that 
$\betrag{f(C) - f_\infty} > \varepsilon$.
Since $f$ is, moreover, almost periodic, there is some $\omega > L - C$,
such that $\betrag{f(c) - f(c + \omega)} < \inv2\varepsilon$ for all $c\in\R$.
In particular, we have with $C + \omega > L$
\bglklein
\inv2\varepsilon > \betrag{f(C) - f(C+\omega)} 
   \geq \betrag{f(C) - f_\infty} - \betrag{f(C+\omega) - f_\infty} 
   > \varepsilon - \inv2\varepsilon = \inv2\varepsilon,
\eglklein\noindent
contradicting our assumption. 
Hence $\supnorm{f - f_\infty} \leq \varepsilon$.
Consequently, $f \ident f_\infty$.

\qed
\epf

\blem
\label{lem:f_cf_0}
Let $f : \R \nach \C$ as well as $c \auf c f(c)$
be almost periodic.

Then $f$ vanishes identically.
\elem
\bpf
Since almost periodic functions are bounded, $c f(c)$ is bounded. Hence, 
$f(c) \gegen 0$ for 
$c \gegen \infty$. Since $f$ is almost periodic, we have
$f \ident 0$ from Lemma \ref{lem:converg_apf}.
\qed
\epf

The definition of almost periodicity can immediately be extended
to functions mapping to Banach spaces $X$: Just replace 
modulus by norm. If $X$ is, e.g., the space of $n \kreuz n$ matrices, then
the almost periodicity of $f : \R \nach X$ is equivalent to the
almost periodicity of all matrix elements of $f$.
Again, constant functions are almost periodic; sums and products of almost
periodic functions are almost periodic as well.

The substitution of $\R$ by an arbitrary locally compact abelian group $G$
is less direct.~\cite{RudinFour} 
Here, one considers the dual group $\Gamma$ of $G$
and then the dual group $\quer G_\Bohr$ of $\Gamma$, where
$\Gamma$ is given the discrete topology.
Via Gelfand transform, $\quer G_\Bohr$ is compact abelian and contains $G$ 
as a dense subset. It is called Bohr compactification of $G$.
Almost periodic functions on $G$ are now the $G$-restrictions of 
continuous functions on $\quer G_\Bohr$.


\begin{thebibliography}{1}

\bibitem{EMS1}
{Dmitri V. Anosov and Vladimir I. Arnold \mbox{}(eds.): {\it Dynamical Systems
  I (Encyclopaedia of Mathematical Sciences 1)}. Springer-Verlag, Berlin,
  1994.}

\bibitem{e90}
{Abhay Ashtekar, Martin Bojowald, and Jerzy Lewandowski: Mathematical structure
  of loop quantum cosmology. {\it Adv. Theor. Math. Phys.} {\bf 9} (2003)
  {233--268}. \\ {\sf e-print:\ gr-qc/0304074}.}

\bibitem{u1}
{Abhay Ashtekar and Jerzy Lewandowski: Background Independent Quantum Gravity:
  A Status Report. {\it Class. Quant. Grav.} {\bf 21} (2004) {R53--R152}. {\sf
  e-print:\ gr-qc/0404018}.}

\bibitem{Bohr}
{Harald Bohr: {\it Fastperiodische Funktionen}. Verlag von Julius Springer,
  Berlin, 1932.}

\bibitem{d80}
{Martin Bojowald: Absence of Singularity in Loop Quantum Cosmology. {\it Phys.
  Rev. Lett.} {\bf 86} (2001) {5227--5230}. {\sf e-print:\ gr-qc/0102069}.}

\bibitem{d83}
{Jonathan Engle: Relating loop quantum cosmology to loop quantum gravity:
Symmetric sectors and embeddings.
  {\it Class. Quant. Grav.} {\bf 24} (2007) {5777--5802}. \\
{\sf e-print:\ gr-qc/0701132}.%
\footnote{Previous title: 
On the physical interpretation of states in loop quantum cosmology.} 
}

\bibitem{diss}
{Christian Fleischhack: Mathematische und physikalische Aspekte
  verallgemeinerter Eichfeldtheorien im Ashtekarprogramm (Dissertation).
  Universit{\"a}t Leipzig, 2001.}

\bibitem{d84}
{Tim A. Koslowski: Holonomies of isotropic $SU(2)$ connections on $\R^3$. {\it
  in preparation}.}

\bibitem{RudinFour}
{Walter Rudin: {\it Fourier Analysis on Groups}. John Wiley \& Sons, New York,
  1990.}

\end{thebibliography}
\end{document}